\documentclass[aps,prl,superscriptaddress,epsfigm,amsmath,twocolumn,showpacs]{revtex4}

\usepackage[dvips]{graphicx}

\usepackage{bm}

\begin{document}

\title{Signatures of molecular correlations in the 
few-electron dynamics of coupled quantum dots}

\author{Andrea Bertoni}
\email[e-mail: ]{bertoni.andrea@unimore.it}
\affiliation{CNR-INFM - National Research Center on nano-Structures
and bio-Systems at Surfaces ($S3$), Via Campi 213/A, 41100
Modena, Italy}
\author{Juan I. Climente}
\affiliation{CNR-INFM - National Research Center on nano-Structures
and bio-Systems at Surfaces ($S3$), Via Campi 213/A, 41100
Modena, Italy}
\author{Massimo Rontani}
\affiliation{CNR-INFM - National Research Center on nano-Structures
and bio-Systems at Surfaces ($S3$), Via Campi 213/A, 41100
Modena, Italy}
\author{Guido Goldoni}
\affiliation{CNR-INFM - National Research Center on nano-Structures
and bio-Systems at Surfaces ($S3$), Via Campi 213/A, 41100
Modena, Italy}
\affiliation{Dipartimento di
Fisica, Universit\`a di Modena e Reggio E., 41100 Modena, Italy}
\author{Ulrich Hohenester}
\affiliation{Institut f{\"u}r Physik, Karl-Franzens-Universit{\"a}t
Graz, Universit{\"a}tsplatz 5, 8010 Graz, Austria}


\begin{abstract}
We study the effect of Coulomb interaction on the few-electron
dynamics in coupled semiconductor quantum dots by exact
diagonalization of the few-body Hamiltonian.  The oscillation of
carriers is strongly affected by the number of confined electrons and
by the strength of the interdot correlations.  Single-frequency
oscillations are found for either uncorrelated or highly correlated
states, while multi-frequency oscillations take place in the
intermediate regime.  Moreover, Coulomb interaction renders
few-particle oscillations sensitive to perturbations in spatial
directions other than that of the tunneling, contrary to the
single-particle case.  The inclusion of acoustic phonon scattering
does not modify the carrier dynamics substantially at short times,
but can damp oscillation modes selectively at long times.
\end{abstract}

\pacs{73.21.La, 03.67.Lx, 72.10.Di}

\maketitle

Low-dimensional heterostructures enable direct probing of the time
evolution of carriers.  In particular, charge oscillations between
coupled quantum structures have been measured in a number of different
systems, such as Josephson junctions \cite{PashkinNAT}, quantum wells
\cite{LeoPRL}, or quantum dots \cite{HayashiPRLPE,GormanPRL}. Coupled
quantum dots (CQDs), where the number of confined carriers can be
controlled experimentally, are a most interesting case as they posses
a discrete energy spectrum which stems from the quantum confinement in
all three spatial directions, thus constituting the physical
realization of the particle-in-the-box problem.  Understanding charge
oscillations in these structures is not only of fundamental importance
but also of technological relevance, since control of the coherent
dynamics of charge states is at the base of many proposals for novel
nanoelectronic devices \cite{LentAPL,AstafievAPL,GywatPRB} and quantum
logic gates \cite{FedichkinNT,vanderWielJJAP,hohenesterPRB06}.
This has motivated recent attempts to achieve control of charge 
localization and dynamics in CQD systems by means of microwave
excitations\cite{PettaPRL}, static\cite{GormanPRL} and
time-dependent\cite{Tamborenea,ZhangPLA} electric fields.

Coherent charge oscillations between CQDs have been recently
demonstrated in the single-electron \cite{PettaPRL} and few-electron
\cite{HayashiPRLPE,GormanPRL} regimes.  The origin and behavior of
these oscillations in the former case is well understood: when a
single electron is placed in one of the dots of a CQD system, with no
other electron in it, it oscillates back and forth between the dots
with a frequency given by the energy difference between the bonding
and antibonding ``molecular'' states.  When a larger number of
electrons is present, the system behavior is much less understood.
Recent experiments working in the latter regime exposed a
single-frequency oscillation of the carriers
\cite{HayashiPRLPE,GormanPRL}, which was interpreted in terms of
an effective single-electron picture.  However, in general one would
expect more complicated oscillation patterns owing to the non-trival
density of states of Coulomb-correlated few-body systems
\cite{ZhangPLA,ZhangPRE,Chakra_book,Pawel_book}.

In this work we theoretically investigate charge oscillations in
few-electron CQDs. We show that, as an effect of Coulomb correlations,
the amplitude, period, and shape of these oscillations are strongly
dependent on the number of electrons confined in the structure.
Either single-frequency or multi-frequency charge oscillations occur,
depending on the strength of the correlations between molecular levels
of the CQD.  Additional external magnetic fields are shown to provide
a versatile means for tuning the period of the charge
oscillations. This unique behavior due to the full spatial
quantization is characteristic of quantum dots and clearly differs
from the plasmon oscillations in coupled quantum wells \cite{LeoPRL,RaichevPRB}.
We also investigate the effect of the acoustic phonon bath, and show
that the different strength of electron-phonon interaction for
different few-electron states leads to selective suppression of modes
in the multi-frequency dynamics.

\begin{figure}[b]
\includegraphics[width=\linewidth]{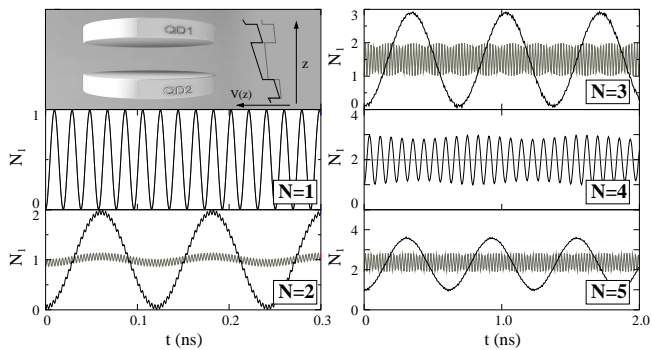}
\caption{
Schematic of the CQD structure under study and
occupation probability in the upper dot as a function
of time for $N$-electron systems.  Gray and black lines
represent the oscillation patterns for weak ($E_z=50$~kV/m)
and strong ($E_z=170$~kV/m) initialization biases, respectively.
The dots have thickness $W=10$~nm, lateral harmonic confinement 
$\hbar \omega_0=3$~meV and the interdot separation is $L_b=8$~nm.
The prominent oscillation period depends mainly on the number of
oscillating electrons (note the two different scales for the time).
}\label{fig1}
\end{figure}

We consider a system of two vertically coupled dots, as depicted in
the first panel of Fig.~\ref{fig1}, populated with $N$ interacting
electrons.  The dots are gated GaAs/AlGaAs heterostructures, as those
built in Ref.~\onlinecite{AustingPB}.  The conduction-band
single-electron states are described within a three-dimensional
envelope function approximation, including an electric field $E_z$
along the $z$ direction \cite{poten_params}. For this cylindrically
symmetric configuration, the single-particle eigenfunctions can be
given the separable form $\psi_{nmgs}({\bf r},\sigma)= \phi_{nm}(x,y)
\varphi_{g}(z)\chi_s(\sigma)$, with $n$ and $m$
radial and angular quantum numbers of the Fock-Darwin state
\cite{Pawel_book}, $g=0,1$ labeling the bonding and antibonding
eigenfunctions associated with the double-well potential, and $s$ the
spin orientation.  In order to compute the few-electron states
exactly, we use a full configuration interaction approach
\cite{RontaniJCP}: the single-particle states are populated with $N$
electrons in all possible ways to construct a basis of Slater
determinants $|\Phi_j\rangle$, where $j$ stands for the set of
many-body quantum numbers.  Then, the three-dimensional
$N$-electron Hamiltonian is diagonalized.  We first assume a closed
system, so that coherent charge oscillations take place.  In the last
part of this work, we briefly investigate the effect of phonons.

Initially, the CQD is subject to an electrostatic bias $E_z$ which
favors localization in the lower dot (QD2).  Then, at time $t=0$, the
bias is removed (non-adiabatically) and the carriers start oscillating
between the two dots.
In order to simulate this process, the configuration interaction
calculation described above is performed twice: first by taking
$E_z>0$, then $E_z=0$.  The computed eigenstates are
$|\tilde{\Psi}_l\rangle= \sum_j
\tilde{c}_{lj} |\tilde{\Phi}_j\rangle$, and
$|\Psi_l\rangle= \sum_j c_{lj} |\Phi_j\rangle$, with and without the
$E_z$ field, respectively.  We take, as the initial state of our
few-particle system, the ground state in the biased condition,
$|\tilde{\Psi}_0\rangle$.  Its time evolution, after the removal of
the bias, can be computed from
\begin{equation}  \label{eq1}
|\tilde{\Psi}_0(t)\rangle = \sum_l e^{-\frac{i}{\hbar}E_l t}
\langle \Psi_l | \tilde{\Psi}_0 \rangle | \Psi_l \rangle.
\end{equation}
Here the sum runs over the $l$ unbiased states, whose energy is $E_l$,
participating in the spectral decomposition of
$|\tilde{\Psi}_0\rangle$.  In order to visualize the charge
oscillations, we evaluate the particle density inside the upper dot
(QD1), $N_1$.  This magnitude has been resolved experimentally using
pump-and-probe techniques \cite{HayashiPRLPE} or single-electron
transistors \cite{GormanPRL}.

In Fig.~\ref{fig1} we illustrate the time-dependent occupancy of QD1
for $N=1,2,3,4,5$.  Gray and black lines represent weak and strong
initialization bias (see caption), respectively, as schematically
reported by the two $V(z)$ profiles in the first panel.  For $N=1$, we
retrieve the expected sinusoidal oscillation, whose frequency is given
by the bonding-antibonding energy separation, $\Delta E_{\mbox{\tiny
BAB}}$.  Both weak and strong biases completely localize the electron
in QD2, thus giving rise to essentially the same curve.  However, as
$N$ increases, the initialization bias starts playing a critical role,
as it determines the number of electrons localized in each dot at time
zero. This, in turn, gives rise to very different oscillation patterns
(compare black and gray curves for $N>1$).

An inspection of the few-electron oscillations in Fig.~\ref{fig1}
shows that they exhibit multiple frequencies with different amplitudes
(most apparent in the $N=2$ panel). This is a signature of Coulomb
interaction in the interdot dynamics, which can be understood from the
coefficients of the spectral decomposition of the biased state,
$\langle \Psi_l | \tilde{\Psi}_0 \rangle$: only the unbiased states
which have a finite overlap will contribute to the time evolution in
Eq.~(\ref{eq1}).  
In the single-electron case the in-plane and vertical degrees of
freedom decouple.  The initial state is localized in QD2 and results
to be the linear combination of one bonding and one antibonding
unbiased states.  This is because the in-plane $xy$ component of the
wave function is the same with and without bias, while, in $z$
direction, the ground biased state is given by $\varphi_{0}(z)+
\varphi_{1}(z)$.  Therefore, only these two states contribute to the
spectral decomposition and the oscillation frequency is given by their
energy difference.
By contrast, in the few-electron case Coulomb interaction mixes the
radial and vertical degrees of freedom, so that the in-plane parts of
the wave functions with and without $E_z$ are no longer identical.  As
a result, the spectral decomposition may involve several pairs of
bonding-antibonding states with different radial wave functions.  Each
pair contributes to the time evolution with its own frequency, given
by its bonding-antibonding energy splitting, and an amplitude that is
proportional to its spectral decomposition weight.

\begin{figure}
\includegraphics[width=\linewidth]{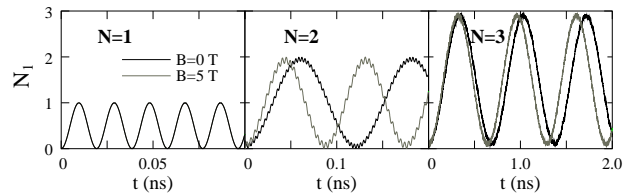}
\caption{
Occupation probability in the upper dot as a function of time for
$N$-electron systems. Gray and black lines represent the oscillation
patterns with ($B=5$ T) and without ($B=0$ T) magnetic field,
respectively.  The CQD structure and biases $E_z$ are the same as in
Fig.~\ref{fig1}.  }\label{fig2}
\end{figure}

The mixing of radial and vertical degrees of freedom due to Coulomb
interaction renders charge oscillations sensitive to perturbations in
the $xy$ plane, even though they do not affect the single-particle
tunneling.
This is shown in Fig.~\ref{fig2},
where we compare single- and few-electron oscillations in the presence
and absence of a magnetic field applied along $z$.  While the $N=1$
oscillation is insensitive to the field, the frequencies of the $N=2$
and $N=3$ ones are increased.  This signature of electron-electron
interaction could be used in experiments to distinguish between the
dynamics of independent and correlated electrons.  It also offers a
unique way to modulate the few-particle oscillation frequency.

\begin{figure}
\includegraphics[width=\linewidth]{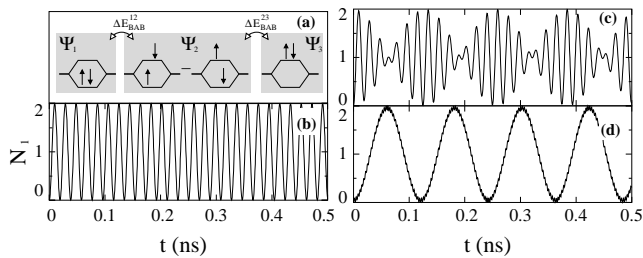}
\caption{
(a) Main unbiased states contributing to the 
spectral decomposition of the ground state in the $N=2$ CQD of
Fig.~\ref{fig1} at $E_z=150$ kV/m. Occupation probability of the upper
dot in systems with: (b) null ($\epsilon^*=\infty$), (c) partially
quenched ($\epsilon^*=10^3$), and (d) regular ($\epsilon^*=12.9$)
Coulomb interaction. 
}\label{fig3}
\end{figure}

To gain further insight into the role of Coulomb interaction, in
Fig.~\ref{fig3}
we compare the oscillation patterns in a CQD with no interaction
(panel b), artificially quenched interaction (panel c), and regular
interaction (panel d).  In all cases, the electric field initializes
both electrons in QD2, and yet major changes take place as the Coulomb
intensity is varied.  Again, the changes can be understood by
analyzing the main unbiased states contributing to the spectral
decomposition of the biased state.  These are the $s$-shell singlets,
$\Psi_1$, $\Psi_2$ and $\Psi_3$, represented in Fig.~\ref{fig3}(a) by
their dominating electronic configuration.  In the schematic
representation, electrons are indicated by arrows, whose direction is
the spin state, and can occupy the single-particle bonding (lower
horizontal line) or anti-bonding (upper horizontal line) states.
For vanishing Coulomb interaction, $\tilde{\Psi}_0$ is composed with
equal weight of the two bonding-antibonding pairs $(\Psi_1+\Psi_2)$
and $(\Psi_2+\Psi_3)$.  The frequency arising from the two pairs is
identical, $\Delta E^{12}_{\mbox{\tiny BAB}}=\Delta
E^{23}_{\mbox{\tiny BAB}}=E_t$ ($E_t$ is the single-particle tunneling
energy), and the oscillation shown in Fig.~\ref{fig3}(b) is
reminiscent of a single-particle one.  Switching on a weak Coulomb
perturbation [Fig.~\ref{fig3}(c)] introduces a small departure from
this limit, here $\Delta E^{12}_{\mbox{\tiny BAB}}=0.24$~meV and
$\Delta E^{23}_{\mbox{\tiny BAB}}=0.20$~meV: now the two pairs have
similar amplitudes but different energies. The superposition of the
two harmonic motions gives rise to a beat with modulated amplitude and
frequency.  Finally, for a realistic Coulomb interaction
[Fig.~\ref{fig3}(d)], $\Delta E^{12}_{\mbox{\tiny BAB}}=1.26$~meV and
$\Delta E^{23}_{\mbox{\tiny BAB}}=0.04$~meV, the weights of the two
bonding-antibonding pairs are very different.  As a consequence, there
is one dominating oscillation mode with large amplitude and low
frequency, and another with small amplitude and high frequency.

\begin{figure}
\includegraphics[width=\linewidth]{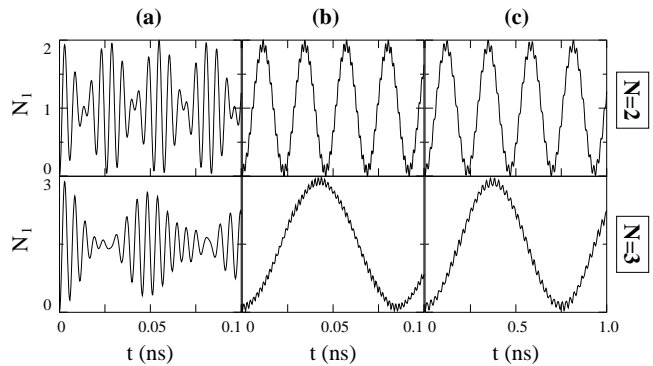}
\caption{
Occupation probability in the upper dot as a function of time for
$N=2$ (upper row) and $N=3$ (lower row).  Here $E_z=500$ kV/m and
$W=10$ nm.  The lateral confinement and interdot barrier thickness in
each column are: (a) $\hbar \omega_0=1$ meV and $L_b=6$ nm; 
(b) $\hbar \omega_0=6$ meV and $L_b=6$ nm;
(c) $\hbar \omega_0=1$ meV and $L_b=10$ nm.
$\epsilon^*=12.9$ in all cases.}\label{fig4}
\end{figure}

The sizable changes of the bonding-antibonding energy splittings in
the presence of Coulomb interaction are an effect of the electronic
correlation between molecular states, often disregarded in previous
studies of multi-particle dynamics in coupled quantum dots\cite{ZhangPLA,ZhangPRE} 
and wells\cite{RaichevPRB}.
The weaker the interdot correlation the stronger the multi-frequency character, 
and vice-versa.  Therefore, one can control the nature of the charge
oscillations by designing CQD structures in the regimes of either weak
or strong interdot correlation with respect to $E_t$.  This is shown
in Fig.~\ref{fig4}, where we compare the $N=2$ and $N=3$ charge
oscillations for different structural parameters, keeping a realistic
value for the Coulomb interaction throughout.  In (a), the interdot
barrier is thin and hence the tunneling energy is large.  The large
splitting between molecular orbitals implies weak interdot
correlation, which leads to an oscillation pattern resembling the
multi-frequency beat of Fig.~\ref{fig3}(c).  In (b), the radial
confinement is increased, which increases the Coulomb repulsion within
the structure. Since the vertical confinement is unchanged, interdot
correlation moves into the strong regime and a quasi-single-frequency
behavior, similar to that of Fig.~\ref{fig3}(d), is retrieved.  In
(c), the barrier is made thicker with respect to (a), so that the
tunneling energy diminishes.  This again enhances the interdot
correlation, leading to the quasi-single-frequency behavior.  We point
out that the small tunneling energy could be the reason for the
single-frequency oscillations reported in
Ref.~\onlinecite{HayashiPRLPE, GormanPRL} experiments.

The electron dynamics in CQDs is severely affected by dissipative
processes \cite{FujisawaJVSTB}.  In what follows, we shall investigate
which novel effects may appear in the few-electron charge oscillations
due to the interaction with the $20$~mK acoustic (deformation-potential and
piezoelectric) phonon bath. 
We calculate the transition rate $\gamma_{li}$ between the correlated
states $\vert \Psi_l \rangle$ and $\vert \Psi_i \rangle$ according to
Ref.~\cite{ClimentePRB} and employ the Pauli master equation to
evaluate the system's time evolution:
$ \frac{d \rho_{ij}}{dt}=
\frac{i}{\hbar} (E_{j} - E_{i}) \rho_{ij} - \sum_l \frac{\gamma_{lj} +
\gamma_{li}}{2} \rho_{ij} + \delta_{ij} \sum_l \gamma_{il} \rho_{ll}
. $
We stress that here $\rho_{ij}$ is an element of the density matrix
corresponding to the (unbiased) few-particle states, with energies
$E_i$ and $E_j$.

\begin{figure}
\includegraphics[width=\linewidth]{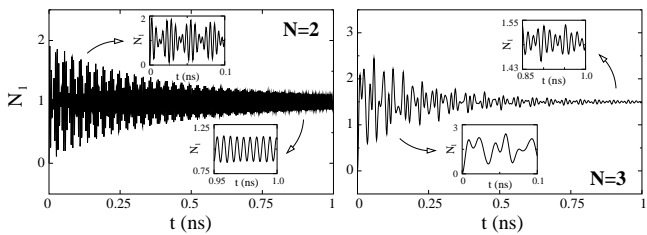}
\caption{\label{fig5}
Occupation probability as a function of time in a 
$N=2$ and $N=3$ CQD subject to phonon interaction. 
The insets zoom-in at short and long times. Note
that the $N=2$ case evolves into a 
single-frequency oscillation.
The CQDs have $\hbar \omega_0$, $W=10$ nm, 
$L_b=6$ nm ($N=2$) and $L_b=8$ nm ($N=3$), and are
initialized with a bias $E_z=500$~kV/m.
}
\end{figure}

Figure~\ref{fig5} shows the resulting charge oscillations for a CQD
with $N=2$ (left panel) and $N=3$ (right panel).  In both cases the
oscillation amplitude is clearly damped by the phonons.  However,
since the transition rate is different for each couple of initial and
final states, the different modes contributing to the multi-frequency
oscillations are damped at different rates.  As a consequence, in some
CQD structures all but one mode are quickly suppressed, and the
initially multi-frequency oscillation turns into a single-frequency
one at later times (compare the insets in the $N=2$ panel for short
and long $t$).  For $N>2$, a higher number of correlated states
participate in the spectral decomposition of the biased state, so that
more oscillation modes show up in the charge oscillation. As a result,
it is difficult to find conditions where only one mode survives the
phonon damping (see insets of $N=3$ panel). Therefore, this is
unlikely to be responsible for the single-frequency oscillations
observed in Refs.~\onlinecite{HayashiPRLPE,GormanPRL}.

In summary, we have shown that the few-electron dynamics of CQDs is
strongly affected by interdot electronic correlations. Drastic changes
in the oscillation pattern, from simple sinusoidal to complicated
beats, take place depending on the correlation strength. 
Electronic interaction further renders few-electron dynamics very
sensitive to perturbations in all directions of the space, and not
only in that of the tunneling. Upon inclusion of phonon damping, 
few-electron charge oscillations may experience an additional effect, 
namely a selective suppression of frequency modes. This is however 
unlikely to be responsible for the single-particle-like oscillations 
reported in early experiments \cite{HayashiPRLPE,GormanPRL}, which 
can be understood as an effect of strong molecular correlation, due 
to the small tunneling energy.
\begin{acknowledgments}
We thank E.~Molinari and F.~Troiani for most helpful discussions.
This work has been supported by projects FIRB no.RBIN04EY74, and
INFM-Cineca Calcolo Parallelo 2007.  J.I.C. acknowledges support from
Marie Curie IEF project MEIF-CT-2006-023797.
\end{acknowledgments}
\vspace{-5mm}

\end{document}